\documentclass[12pt]{article}
\usepackage{graphicx}
\usepackage[bookmarksnumbered,colorlinks,plainpages]{hyperref}

\begin{document}

\title{How to fit the degree distribution of the air network?}

\author{W. Li$^{\dag\ddag}$, Q.A. Wang$^{\dag}$, L. Nivanen$^{\dag}$, and A. Le M\'ehaut\'e$^{\dag}$\\
$^{\dag}$Institut Sup\'erieur des Mat\'eriaux du Mans, \\
44, Avenue F.A. Bartholdi, 72000 Le Mans, France\\
$^{\ddag}$Institute of Particle Physics, \\Hua-Zhong Normal
University, Wuhan 430079, P.R. China}

\date{}

\maketitle

\begin{abstract}
{\em We investigate three different approaches for fitting the
degree distributions of China-, US- and the composite China+US air
network, in order to reveal the nature of such distributions and
the potential theoretical background on which they are based. Our
first approach is the fitting with q-statistics probability
distribution, done separately in two regimes. This yields
acceptable outcomes but generates two sets of fitting parameters.
The second approach is an entire fitting to all data points with
the formula proposed by Tsallis et al. So far, this trial is not
able to produce consistent results. In the third approach, we fit
the data with two composite distributions which may lack
theoretical support for the moment.}
\end{abstract}


\section{Introduction}
Studying properties of various types of networks has recently been
a trend in many different research fields \cite{WS, Watts,
MovieActor, ChemicalReaction, FoodWebs, Ferrer-Janssen-Sole,
Albert-Barabasi-2, Albert-Jeong-Barabasi, Jeong, Yook, Newman1,
Liljeros, Ferrer-Sole}. Nature provides not only large-sized
(herein and after the size of a network refers to the number of
nodes within it) networks such as human relationship network,
Internet \cite{Albert-Barabasi-2}, but also small-sized ones such
as air network for a certain country, and food webs
\cite{FoodWebs}, etc. For example, the current US air network is
undoubtedly the largest one of the same category in the world.
Even so, its size \cite{QuoteUSAir} is only 215
\cite{USAirnetwork}, many decades smaller than those of many
artificial networks which can amount to millions. Obviously,
larger networks are more likely to have better statistics than
their smaller counterparts. An explicit difficulty one may
encounter in dealing with small-sized networks is that due to the
size limit, the nature of distributions of some key quantities,
for instance degree distributions, may be unclear. It is simply
hard for one to draw any credible conclusions because of the
presence of statistical errors. Hence faced with such situations
it is better to resort to possible solutions rather than ascribe
all the faults to the poor statistics.

Loosely speaking, the degree distribution informs us the tendency
of how the whole network is organized. In other words, from the
degree distribution one can have a rough idea of what the network
topology may look like. For instance, if the distribution is
Poisson-like or Gaussian-like, we may conjecture that nodes are
connected more in a random way, or that any two nodes in the
network are connected with nearly equal probability without any
pair being more favored. If the degree distribution is of
scale-free type, then there probably exists a few hubs with many
connections whereas many more nodes have very small degrees. It
was assumed that in the scale-free networks the rule of the
so-called "preferential attachment" \cite{Albert-Barabasi-2}
governs the probability that nodes are connected to one another.
Simply put, "preferential attachment" means that during the
formation of scale-free networks, the highly-connected nodes have
greater chances than the sparsely-connected ones to be connected
by other nodes, which is similar to the phenomenon "rich gets
richer".

There is at least one common thing in dealing with random- and
scale-free networks, that is, one can mathematically explain the
origin of their degree distributions. For some types of networks,
their degree distributions may not follow the standard
distributions as we mentioned above or any other well-known ones.
One good example that can enter here is the air network we have
studied \cite{ChinaAirNetwork, USAirnetwork}. We find that the
cumulative degree distributions (to deal with the statistical
errors the cumulative distribution was introduced) of both China-
and US air networks span two distinctive regimes with a
cross-over, similar to double-pareto law \cite{Double_Pareto}. In
this case, it would be very interesting to examine more carefully
the real nature of such distributions \cite{WhiteTsallis}. After
the fingerprints have been identified, one may further check how
air networks come into being.

In this paper, we will present three different approaches to
fitting the degree distribution of China-, US- and China+US air
network. Section 2 is about the fitting based on the probability
distribution of q-statistics, which is done separately in two
regimes. Section 3 deals with an entire fitting to the formula
proposed by Tsallis et al. In section 4 fittings of the data to
two composite distributions are given. But the theoretical origin
of such distributions are not yet found. The last section is a
brief conclusion.

\section{Fitting with Tsallis-Statistics}
Composed of a number of airports and flights, air networks are
endowed the following characteristics: (a) quite limited system
sizes, being a few hundred at most; (b) relatively stationary
structures with respect to both time and space; (c) bi-directional
flights with slightly fluctuating weights (frequency). In the
terminology of network, the degree $k$ of a certain airport means
it has flights with $k$ other airports in the same network. A very
important quantity related is the distribution of $k$, $p(k)$,
usually called degree distribution, which gives the probability of
finding an airport connected with exactly $k$ other airports
within the same network.

China air network contains 128 commercial airports, and for US air
network, the number is 215. Here we also consider a composite air
network which includes the airports both in China and in US. Hence
the composite China+US air network consists of 343 airports.
Besides all the domestic flights of the two original sub-networks,
the newly composed network also includes a few international
flights. Since the number of international flights is much much
smaller than that of domestic ones, the composite network can be
viewed as superposition of two independent sub-networks.

At a glance of the degree distribution of either China air network
or US air network, we would notice that neither of them follows a
power-law in a whole. But cutting the whole curve into two parts
from a certain transition point, we obtain two straight lines on a
logarithmic co-ordinate. This means each single part is a
power-law.

Power-law distributions are ubiquitous in nature, such as Zipf's
Law \cite{ZipfLaw}, size distribution of earthquakes
\cite{Earthquake}, energy distribution of solar flares
\cite{SolarFlares} and so on. A power-law distribution can be
expressed as
\begin{equation}\label{PowerLaw}
p(x)=Cx^{-a},
\end{equation}
\noindent where $C$ is the normalization constant and $a$ is the
exponent of the law. Power-law is also called scale-free
distribution because its shape remains unchanged whatever we
change its scale, whether magnify or decrease. There are some
claimed mechanisms that can generate power-law distributions, for
instance, combinations of exponentials \cite{Miller}, inverses of
quantities \cite{Newman2}, random walks and Yule process
\cite{Yule}. Among the numerous types of such mechanisms there is
one theory called self-organized criticality (SOC) (\cite{Bak}).
In SOC, events occur in the way of avalanches whose sizes can vary
from a few to a million and obey a power-law which can extend to
many decades.

Two main reasons may account for our motivation of choosing the
probability distribution of Tsallis statistics to fit the degree
distributions of air networks. First, the air network is not a
system which can reach the state of equilibrium. Like many other
complex systems, the air network consists of many units, between
which there are complicated interplays (interactions). Such
systems can not be comfortably treated as simple thermodynamical
systems. Second, as we may have known, Tsallis statistics
\cite{Tsallis1} provides a rather natural way from information
consideration to generate power-law distribution. As a potential
generalization of the conventional Shannon information theory and
the concomitant statistics, the probability distribution of
Tsallis statistics can be written as
\begin{equation}\label{TsallisStatistics}
p(x_i)=\frac {1}{Z_q} [1-(1-q)\beta x_i]^{\frac {1} {1-q}},
\end{equation}
where $x_i$ is the value of a certain quantity at the state $i$,
$Z_q=\sum_{x_i}[1-(1-q)\beta x_i]^{\frac {1} {1-q}}$ is the
partition function and $q$ is a positive index. From the
observation of degree distributions of air networks, it is rather
natural and straightforward to use the following fitting
functions,
\begin{equation}\label{FitTsallisTwo}
p(k)=\frac{[1-(1-q_i)\beta_i k]^{\frac {1}
{1-q_i}}}{\sum_{k}[1-(1-q_i)\beta_i k]^{\frac {1} {1-q_i}}}, i=1,2
\end{equation}
\noindent where $q_1,\beta_1$ and $q_2, \beta_2$ are the
parameters for the small $k$ and large $k$ regime, respectively.

Our fitting using Eq. (\ref{FitTsallisTwo}) and the method of
least squares has been given in Fig. 1, where the top-, middle-
and bottom panel are for China-, US- and China+US air network,
respectively. Their respective fitting parameter sets $(\beta_1,
\beta_2)$ are (0.46$\pm$0.005, 2.85$\pm$0.01), (0.67$\pm$0.003,
3.34$\pm$0.02) and (0.61$\pm$0.003, 4.05$\pm$0.02).
Correspondingly, values of $(q_1, q_2)$ are (3.16$\pm$0.01,
1.35$\pm$0.007), (2.49$\pm$0.01, 1.30$\pm$0.005) and
(2.65$\pm$0.01, 1.25$\pm$0.006). We can see that the three
different systems have different $q$'s. Also, the slopes of the
two separate lines (logarithmic) for China air network and US air
network are nearly consistent with what we obtained in Refs.
\cite{ChinaAirNetwork} and \cite{USAirnetwork}.

\section{An ambitious fitting approach}
In this part, a more ambitious though tougher fitting approach
will be given, adopting the method suggested by Tsallis et al
\cite{TsallisGeorgeMendes}. According to Ref.
\cite{TsallisGeorgeMendes} we assume that, the solution of the
following equation has the tendency to describe the different
behaviors of degree distributions at two separate regimes which
meet at a transition point,
\begin{equation}\label{WholeDistributionAirNetwork}
\frac {d p(k)}{dk}=-\mu_r p^r(k)-(\lambda_q-\mu_r)p^q(k),
\end{equation}
\noindent with $r \le q$. Here $\mu_r$, $\lambda_q$, $q$ and $r$
are four parameters which can be determined through normalization
of the degree distribution $p(k)$. It was claimed that $1/(1-q)$
and $1/(1-r)$ represent the slopes of the two different parts of
the degree distributions (logarithmic) respectively. One specific
choice is $r=1$ and $q > 1$. But apparently such an option is not
feasible since the slope of the second line segment is not
infinity. What we can only resort to is the more generic case $1 <
r < q$, and thereby the solution of Eq.
(\ref{WholeDistributionAirNetwork}) satisfies the following
integral equation \cite{TsallisGeorgeMendes}
\begin{equation}\label{IntegralFitting}
k=\int_{p(k)}^1 \frac {dx} {\mu_r x^r+(\lambda_q-\mu_r)x^q}.
\end{equation}
\noindent Further calculation of Eq. (\ref{IntegralFitting}) using
Mathematica leads to \cite{TsallisGeorgeMendes}
\begin{eqnarray}
\label{HypergeomFit} \nonumber
  k&=& \frac{1}{\mu_r} \{ \frac{p^{1-r}(k)-1}{r-1}-\frac{\lambda_q/\mu_r-1}{1+q-2r} \\ \nonumber
   &&\times [H(1;q-2r,q-r,(\lambda_q/\mu_r-1)) \\
   && -H(p(k);q-2r,q-r,(\lambda_q/\mu_r-1))]\},
\end{eqnarray}
\noindent where $H(x;a,b,c)=x^{1+a}F(\frac
{1+a}{b},1;\frac{1+a+b}{c};-x^bc)$, with $F$ being the standard
hypergeometric function.

After the above preparations in the theoretical aspects, what is
left seems simply fitting the data to appropriate equations.
However, the actual fitting procedure was not at all smooth and
many technic details have to be resolved. Now we have at least
three options in choosing which equation is used to fit the data.
Which one, among Eqs. (\ref{WholeDistributionAirNetwork}),
(\ref{IntegralFitting}), and (\ref{HypergeomFit}), is more
suitable? Let us start from Eq.
(\ref{WholeDistributionAirNetwork}). Initially one needs to
compute the set of first derivatives $dp(k)/dk$ from the data,
which is rather trivial. Then one can readily obtains the values
of the four parameters by means of least squares. The disadvantage
is that due to the small number of data points available, it is
hard to establish a solid relationship between $dp(k)/dk$ and
$p(k)$, and the existence of such arbitrariness may greatly hamper
the exactness of the parameters. That is, the fitting error could
be rather large so that the fitting is not ideal. The advantage is
the simple, straightforward performance. The second choice of
fitting, by using Eq. (\ref{IntegralFitting}), is mainly affected
by the problem of singularity. More precisely, certain
combinations of values of parameters will cause the integral
kernel on the right-hand side of Eq. (\ref{IntegralFitting}) to
diverge. This kind of difficulty could be avoided by restricting
the range of parameters. But how could we be sure that the fitting
has not been affected by doing so? Lastly, if Eq
(\ref{HypergeomFit}) is employed for fitting, the biggest
challenge will be dealing with the hypergeometric functions which
are infinite series. Apparently we are unable to calculate the sum
of infinite series unless we can judge that it converges. Even if
you know the sum is limited, you are still faced with problems
such as how to make a reasonable cut-off on the series.

So far, our fittings using the method of least squares and the
equations in this section are not able to provide satisfiable
outcomes. One of our fitting trials on China air network has been
shown in Fig. 2. It can be seen that the fitted curve can not
match most of the data points--only the tail is well fitted, and
the fitting of other parts is rather poor. Other combinations of
parameters have also been tried but given no better results. If
both the first few points and the tail are included, the
intermediate part will deviate from the curve a lot. It is simply
not easy to compromise all different parts.

Requested by us, Borges tried in a different but less standard way
to do the same fitting with our data. Initially he followed the
method in Ref. \cite{TsallisGeorgeMendes} to estimate the values
of $\mu_r$, $\lambda_q$, $q$ and $r$ directly from the curves
depicting the original data. Then from Eq. (\ref{IntegralFitting})
he calculated the values of $k$ as he treated the values of $p(k)$
as inputs. His "fitting" results have been shown in Figs. 3, 4 and
5. But there is still a problem in his fitting. As we can see from
Figs. 3, 4 and 5, the fitting values of $r$ for the three
different air networks are all 0.6, less than 1. But $r \ge 1$ is
required by the method he used. Also, if we check the curves of
degree distributions, we notice that the slopes of the second
parts are apparently larger than 1. If the claim by Ref.
\cite{TsallisGeorgeMendes} that $1/(1-r)$ is the slope of the
second part is correct, we deduce that $r$ should be larger than
1. How should we explain the discrepancy between the theoretical
background and his fitting?

\section{Fitting approaches using composite distributions}
As a matter of fact, Eq. (\ref{WholeDistributionAirNetwork}) is a
sort of composition of two different power laws in the form of
differential equations. Inspired by this approach, we tried to
compose appropriate distributions which could match the entire
curves of the degree distributions. The first candidate coming
into our mind is expressed as
\begin{equation}\label{PowerLawComposite}
p(k)=ak^{r_1}+bk^{r_2},
\end{equation}
\noindent where the parameters $a$, $r_1$, $b$ and $r_2$ can be determined from the
normalization. Initially we intend to combine two power-laws with negative
exponents, that is $r_1 < 0$ and $r_2 < 0$. But the best fitting with Eq.
(\ref{PowerLawComposite}) to the data does not indicate that both are less than 0.
The real thing is, if $r_1$ is less than 0, then $r_2$ will be greater than 0.
Otherwise, if $r_1 > 0$ is found, then $r_2 <0$ is obtained. Our fitting using the
method of least squares for the three air networks have been shown in Figs. 6, 7 and
8. From the three figures we notice that the heads are all well fitted whereas the
transition parts and the tails do not cooperate. If we check the values of the
fitting parameters, we will find that the exponents, that is, -0.2633, -0.4046, and
-0.2862 are close to the slopes of the first segment lines of log-log degree
distributions for the three air networks, respectively.

Another distribution we can compose is,
\begin{equation}\label{ReverseComposite}
p(k)=\frac{1}{ak^{r_1}+bk^{r_2}}.
\end{equation}
\noindent This relationship came to us just by a mathematical consideration in order
to reproduce two regime distributions after the failure of Eq.
(\ref{PowerLawComposite}) which did not show distinctive transition between the
lower and higher degree parts. Eq. (\ref{ReverseComposite}) has a quite different
behavior from Eq. (\ref{PowerLawComposite}) and shows a distinctive transition
``knee'' like the observed data. It appears from Figs. 9, 10 and 11, that the data
is pretty well matched with Eq. (\ref{ReverseComposite}), by means of least squares,
for all the three networks. In addition, the values of $-r1$ and $-r2$ nearly
represent the slopes of the two separate line segments of the degree distributions.
Take China air network as examples. The fitting parameters therein are
$a=2.022e-8,r_1=5.001,b=0.9376,$ and $r_2=0.3608$. When $0 < k <k_c$ ($k_c$ is the
degree of the transition point which can be determined through
$ak_c^{r_1}=bk_c^{r_2}$), there will be $bk^{r_2} \gg ak^{r_1} $), and hence $p(k)
\sim k^{-r_2}$. When $k > k_c$, then $ak^{r_1} \gg bk^{r_2}$, and hence $p(k) \sim
k^{-r_1}$.

\section{Conclusions}
In summary, we have fitted the degree distributions of air network
in China, in US and in China+US in several different ways. The
first approach leads to two-regime power-laws, each of which can
be well described by probability distribution of Tsallis
statistics. However, the fitting generates two $q$'s, one for
small degree region and another for large degree region. How could
we explain that the value of $q$ is different even within the same
system? Why should we divide the whole distribution into two
parts? This man-made separation is apparently arbitrary. Could we
thus believe that there exists different hierarchies in the
organization of the air networks? As pointed out by \cite{Alain},
should small airports stay in a group where the law is based on a
certain reference, while the larger airports stay in another one
where the law is based on a different reference? The observation
is not sufficient for us to arrive at the conclusion that air
network is an non-extensive system. The second type of fitting
approach, also based on Tsallis statistics but having a more
generic form, provides the possibility of an entire fitting to all
the data points. But so far, we are unable to come up with any
consistent results by using the method. The third type of fitting
approach can help to find some distributions well matched with the
data but lacking theoretical background. That is, how can we
derive such distributions from the first principle or at least in
a reasonable way?

\section*{Acknowledgement}
Authors would like to thank C. Tsallis and E. Borges for their
fruitful discussions when this work was done. This work is
supported in part by National Natural Science Foundation of China
and the R\'egion des Pays de la Loire of France under Grant $N^o$
04-0472-0.

\newpage

\pagestyle{empty}
\begin{figure}[hb] \label{f1}
\includegraphics[width=8cm,height=18cm]{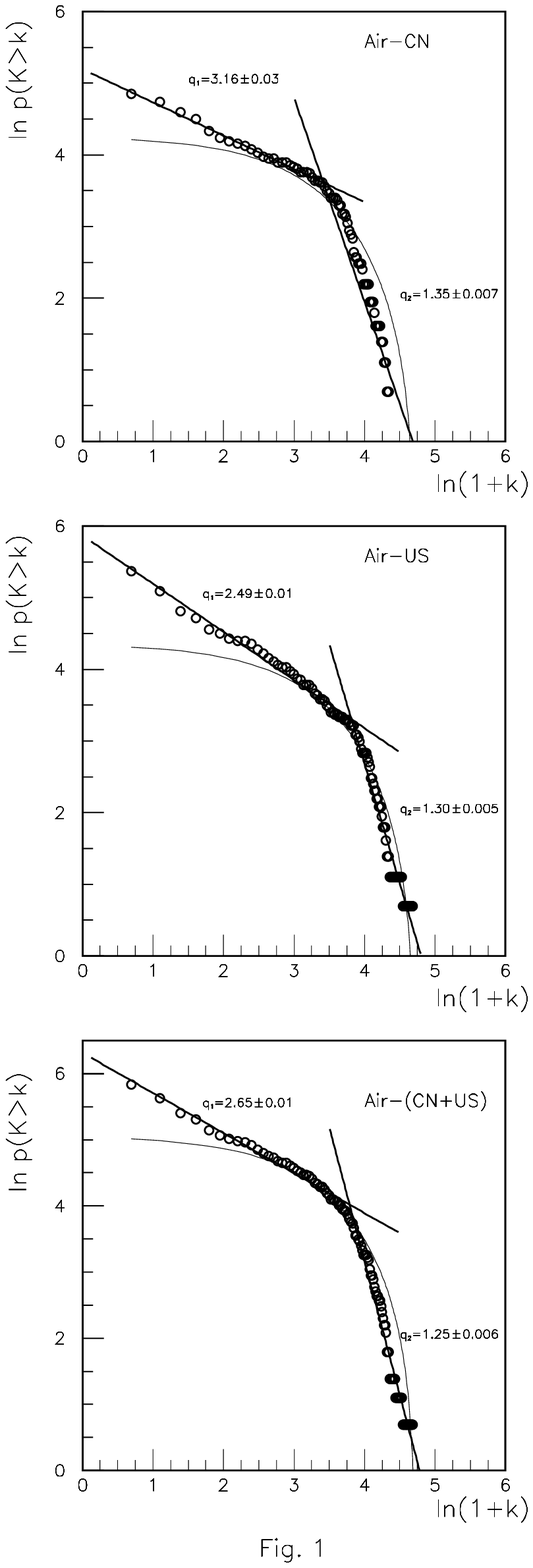}
\caption{Degree distributions (circles) of China air network (top
panel), US air network (middle panel), and China+US air network
(bottom panel). The straight lines are least squares fittings
with the probability distribution of q-statistics given by
Eq.(\ref{FitTsallisTwo}). In order to compare the observed
two-regime distribution with exponential law, the latter is also
drawn in the figure by using curved lines.}
\end{figure}

\begin{figure}[h] \label{f2}
\includegraphics[width=16cm,height=20cm]{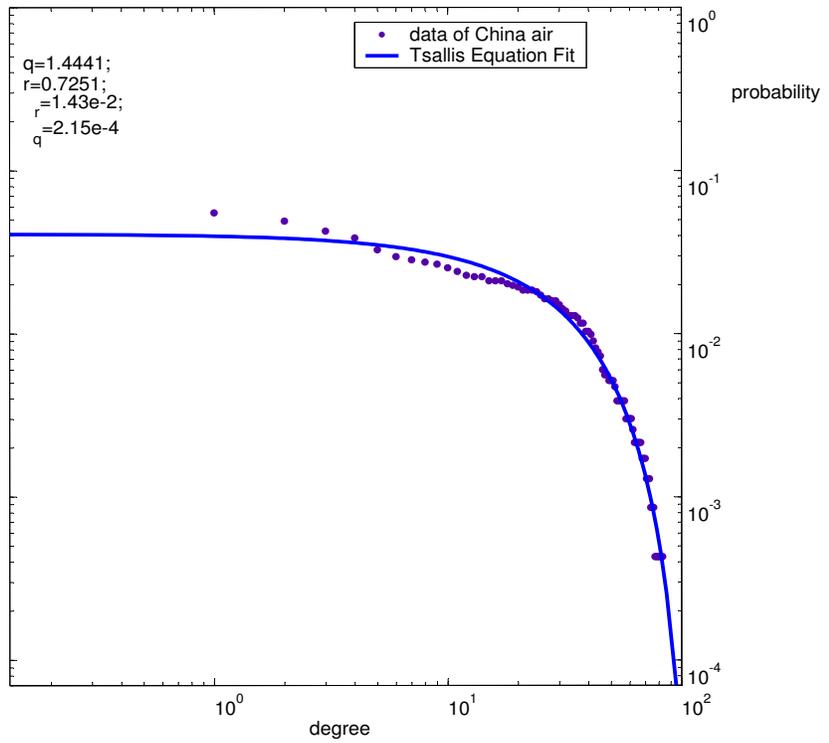}
\caption{Degree distribution (blue points) of China air network.
The blue line is least squares fitting with Eq.
(\ref{HypergeomFit}) where the first 500 series of the
hypergeometric function was taken.}
\end{figure}

\begin{figure}[hb] \label{f3}
\includegraphics[width=19cm,height=19cm]{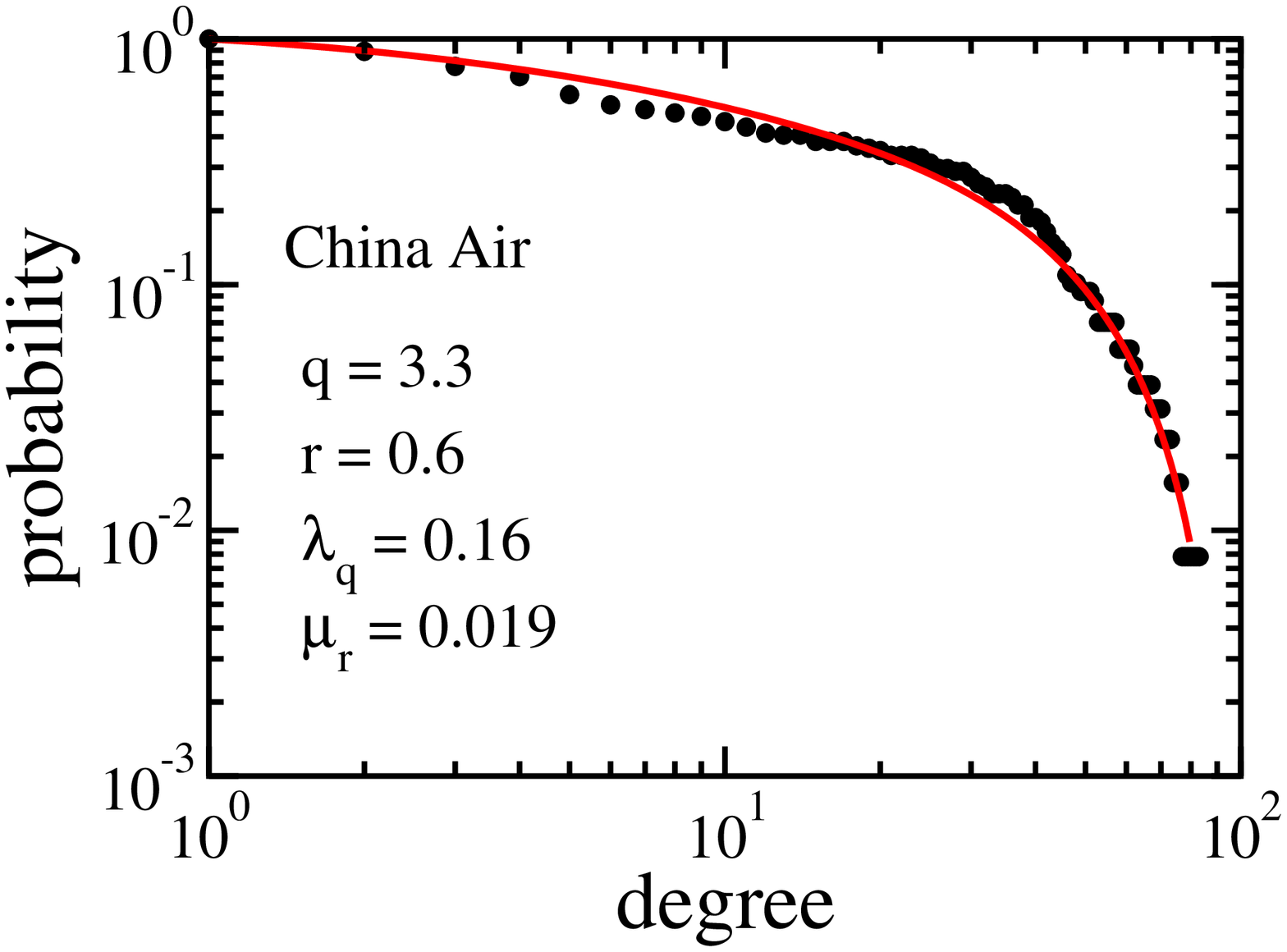}
\caption{Degree distribution (black points) of China air network.
The red line represents the fitting with Eq.
(\ref{IntegralFitting}) where the four parameters $\mu_r$,
$\lambda_q$, $q$ and $r$ were estimated directly from the data
points.}
\end{figure}

\begin{figure}[hb] \label{f4}
\includegraphics[width=19cm,height=19cm]{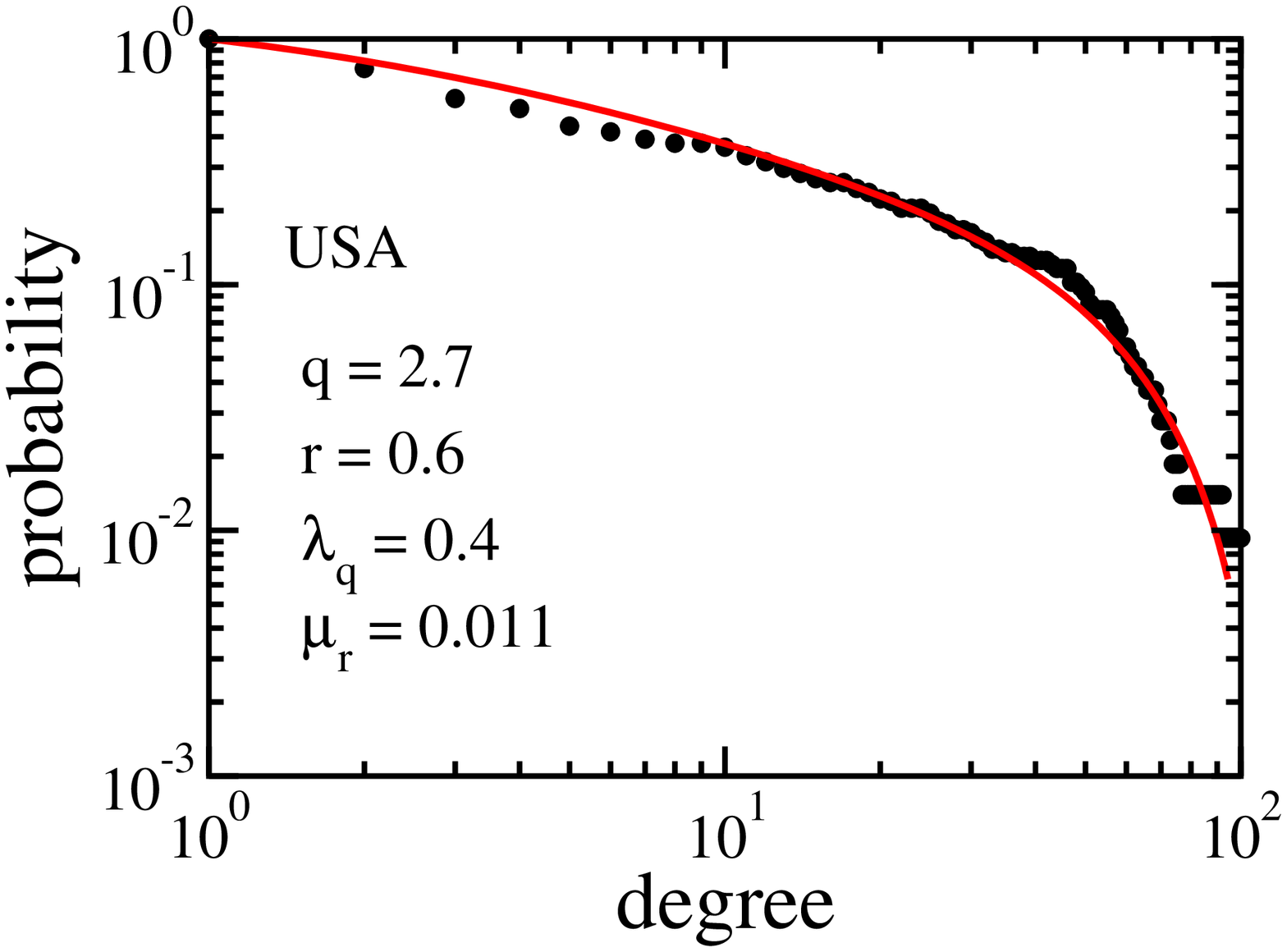}
\caption{Degree distribution (black points) of US air network. The
red line is the fitting with Eq. (\ref{IntegralFitting}) where the
four parameters $\mu_r$, $\lambda_q$, $q$ and $r$ were estimated
directly from the data points.}
\end{figure}

\begin{figure}[hb] \label{f5}
\includegraphics[width=19cm,height=19cm]{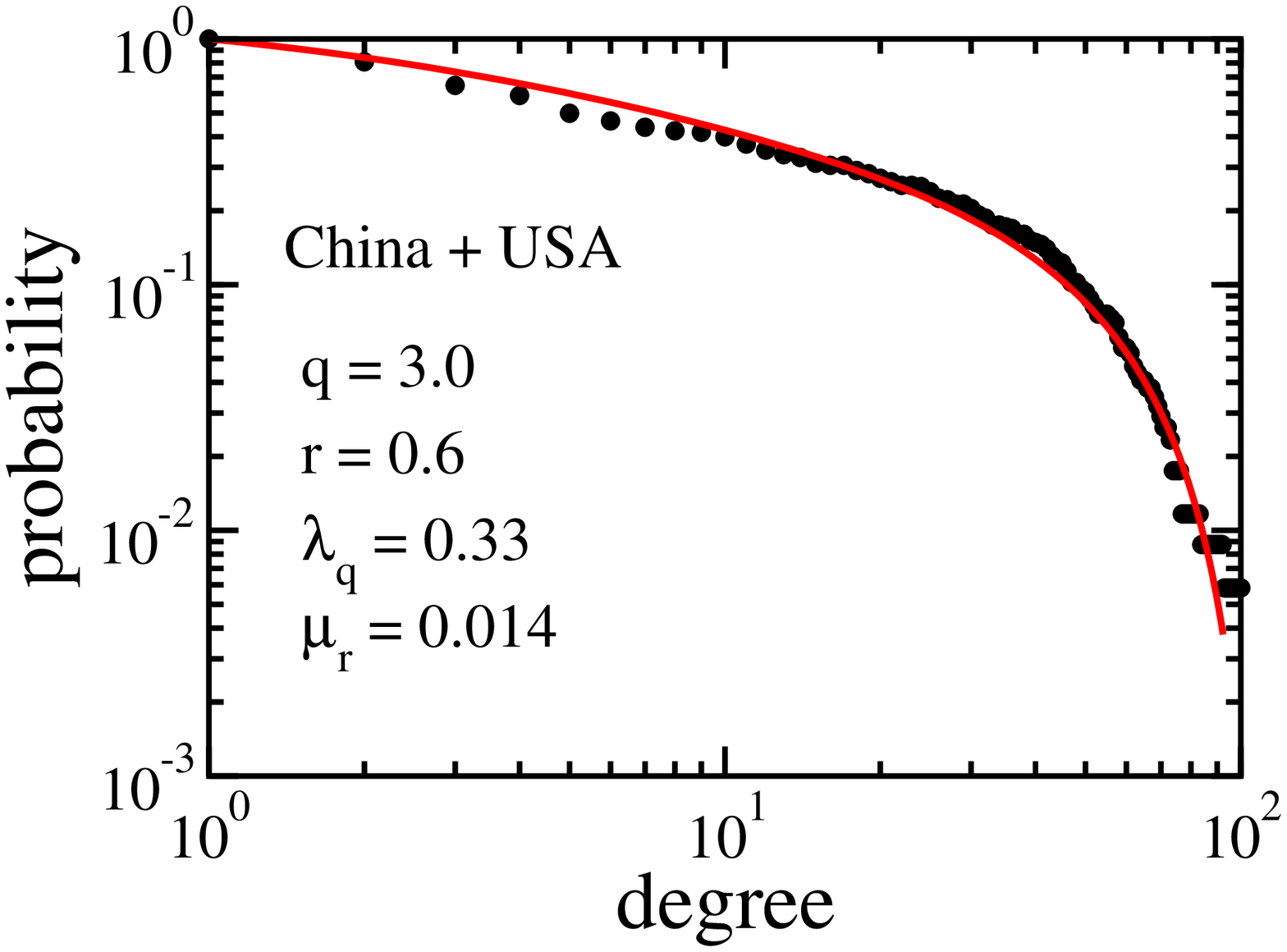}
\caption{Degree distribution (black points) of China+US air
network. The red line shows the fitting with Eq.
(\ref{IntegralFitting}) where the four parameters $\mu_r$,
$\lambda_q$, $q$ and $r$ were estimated directly from the data
points.}
\end{figure}

\begin{figure}[hb] \label{f6}
\includegraphics[width=15cm,height=19cm]{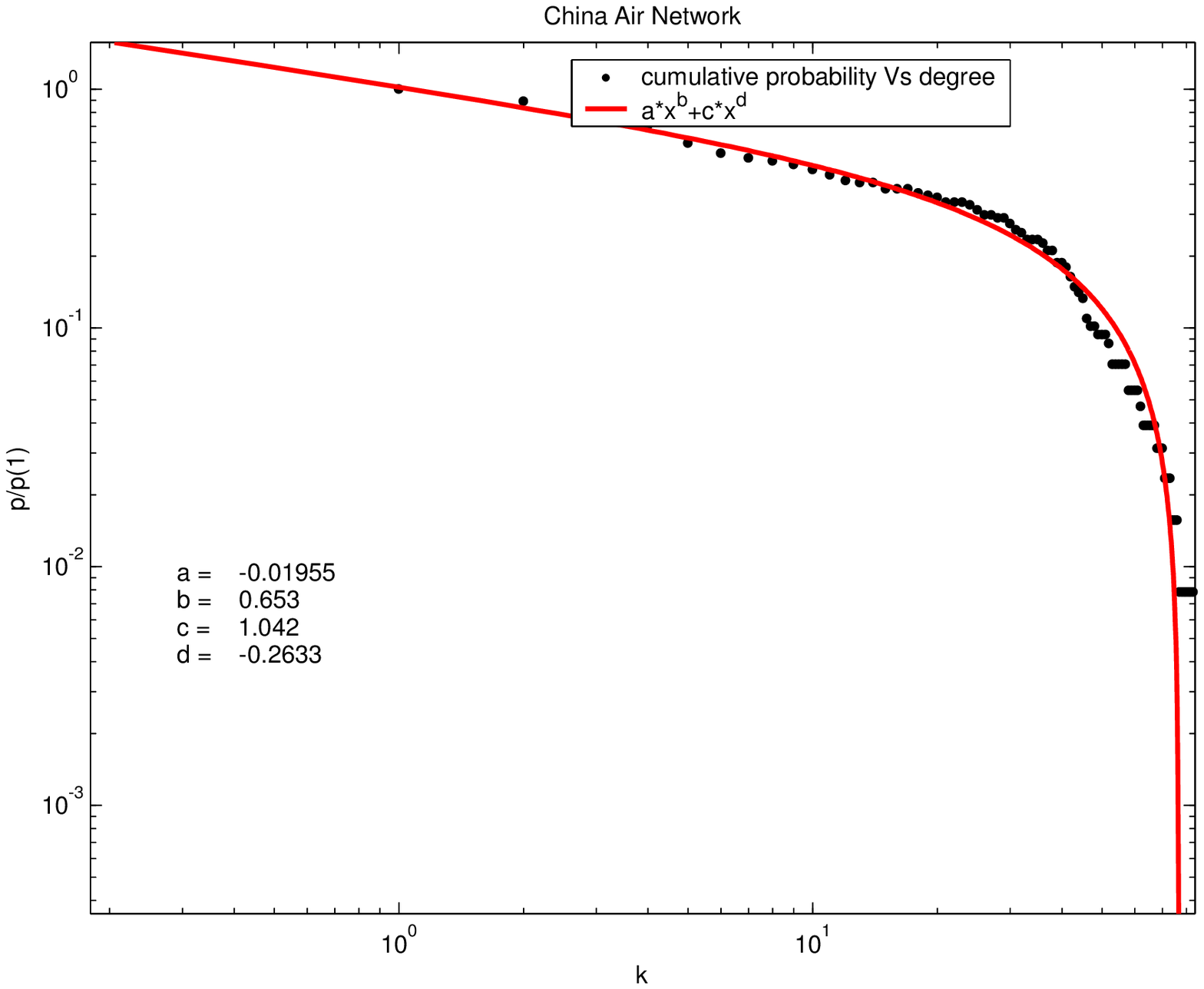}
\caption{Degree distribution (black points) of China air network.
 The red line is the least squares fitting with Eq. (\ref{PowerLawComposite}).}
\end{figure}

\begin{figure}[hb] \label{f7}
\includegraphics[width=15cm,height=19cm]{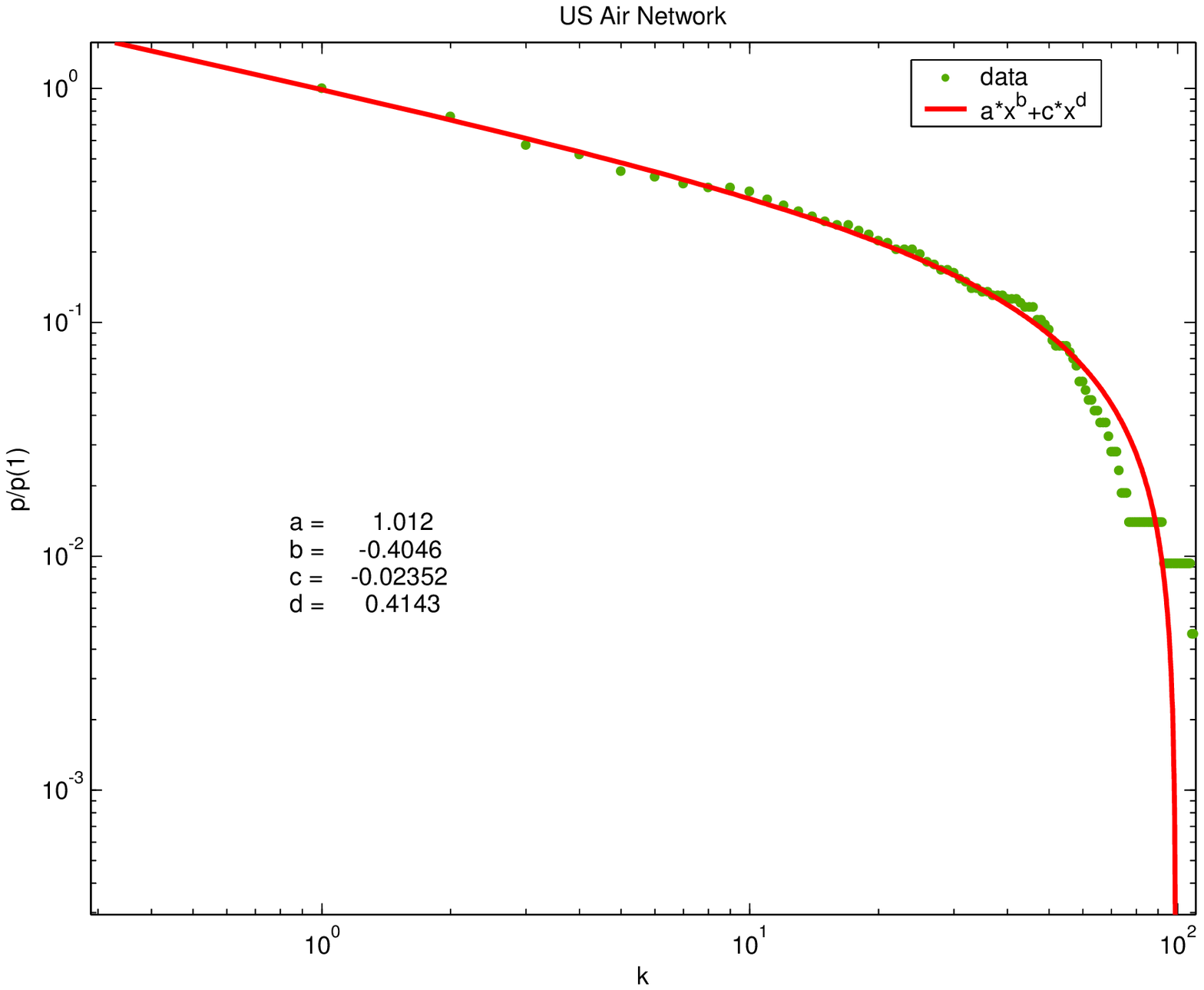}
\caption{Degree distribution (black points) of US air network. The
red line is the least squares fitting with Eq.
(\ref{PowerLawComposite}).}
\end{figure}

\begin{figure}[hb] \label{f8}
\includegraphics[width=15cm,height=19cm]{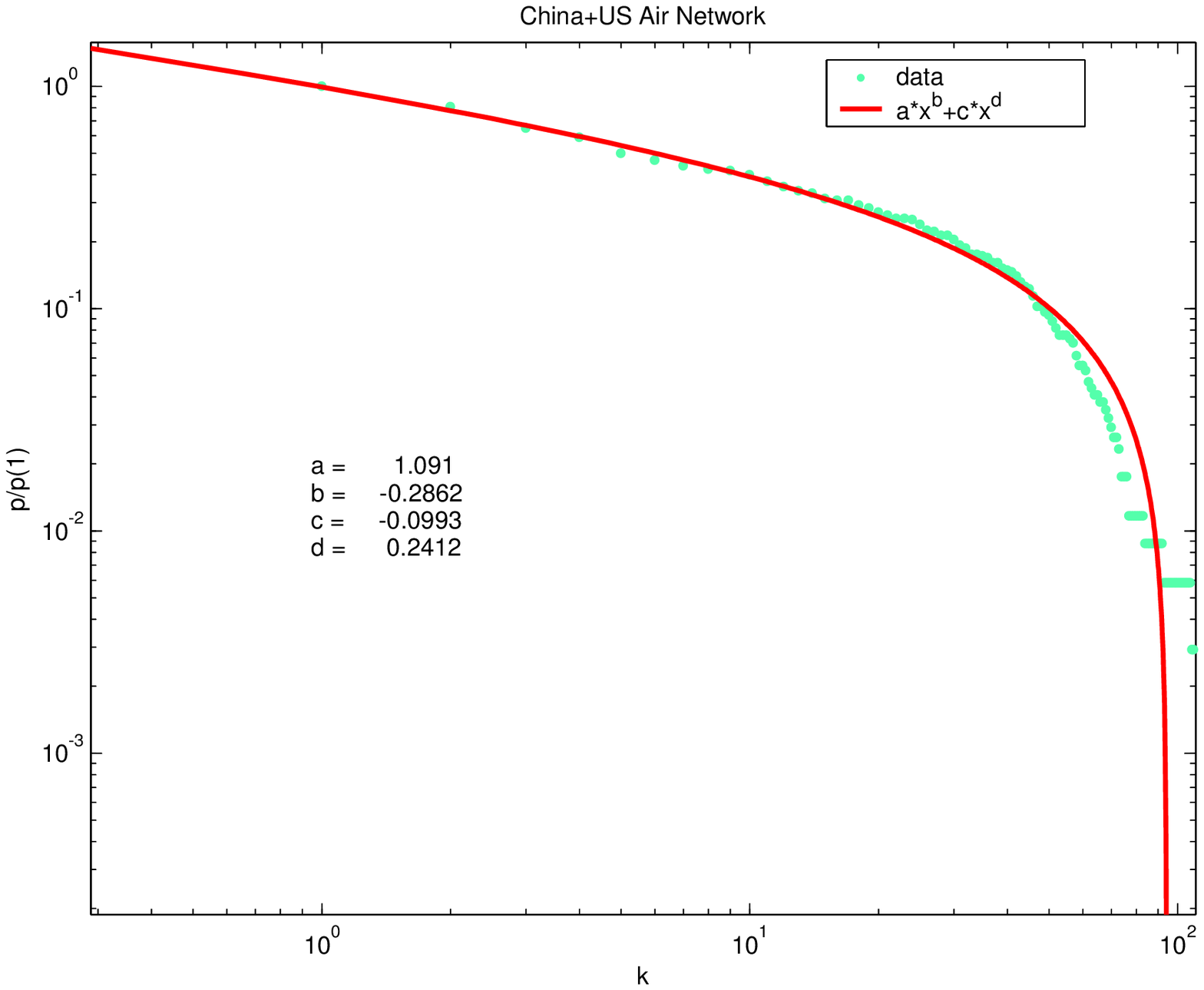}
\caption{Degree distribution (black points) of China+US air
network. The red line is the least squares fitting with Eq.
(\ref{PowerLawComposite}).}
\end{figure}

\begin{figure}[hb] \label{f9}
\includegraphics[width=15cm,height=19cm]{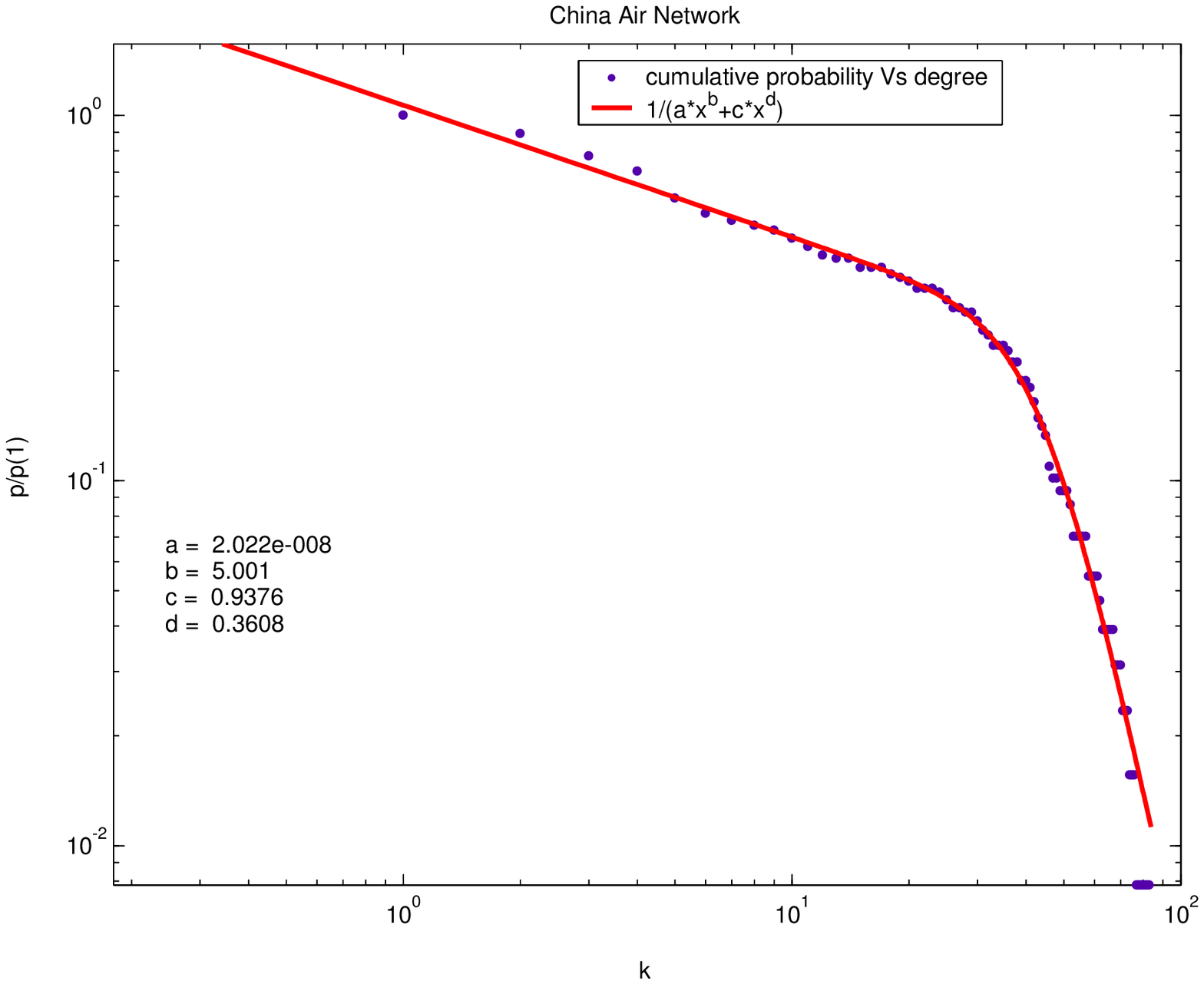}
\caption{Degree distribution (black points) of China air network.
The red line is the least squares fitting with Eq.
(\ref{ReverseComposite}).}
\end{figure}

\begin{figure}[hb] \label{f10}
\includegraphics[width=15cm,height=19cm]{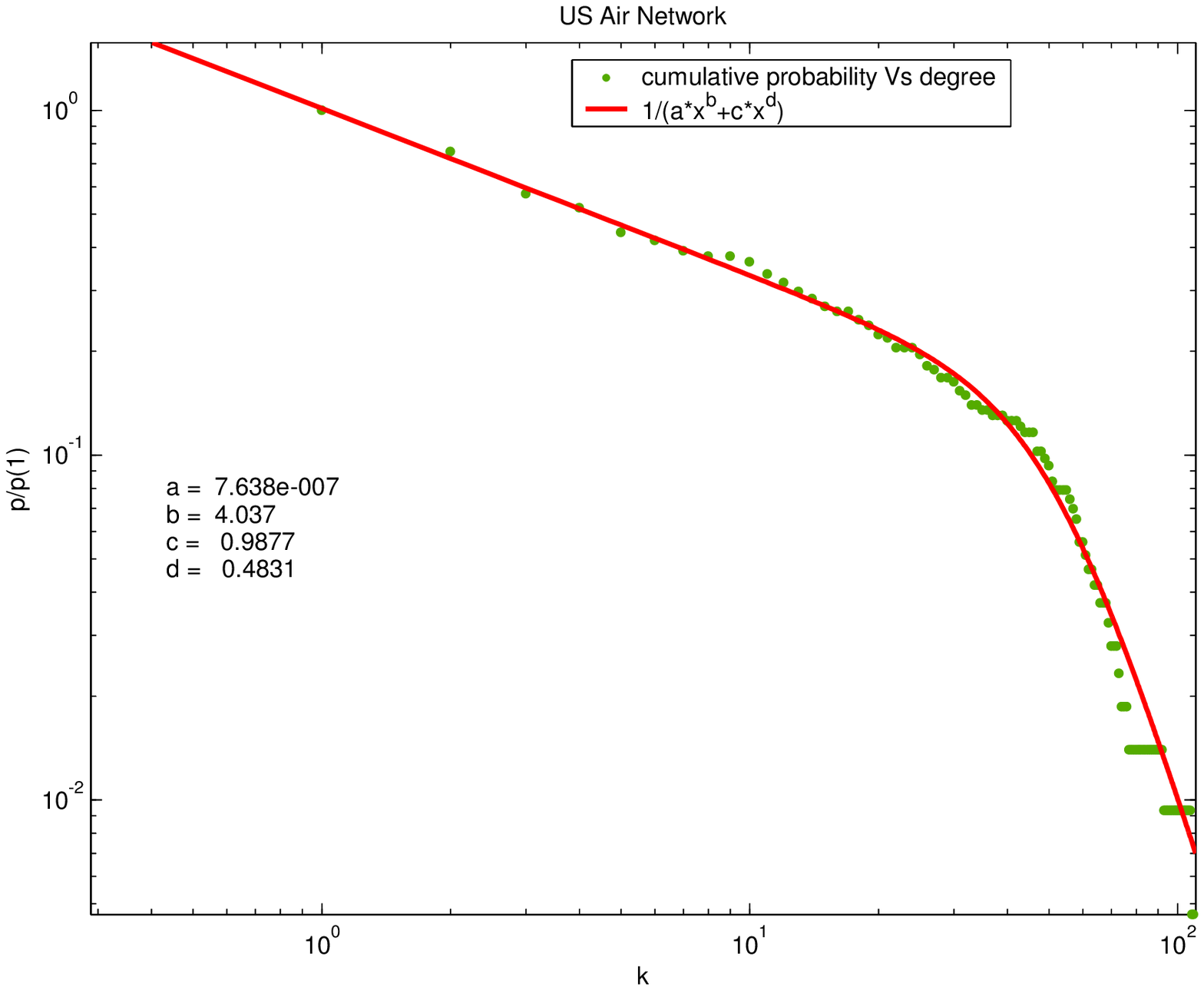}
\caption{Degree distribution (black points) of US air network. The
red line is the least squares fitting with Eq.
(\ref{ReverseComposite}).}
\end{figure}

\begin{figure}[hb] \label{f11}
\includegraphics[width=15cm,height=19cm]{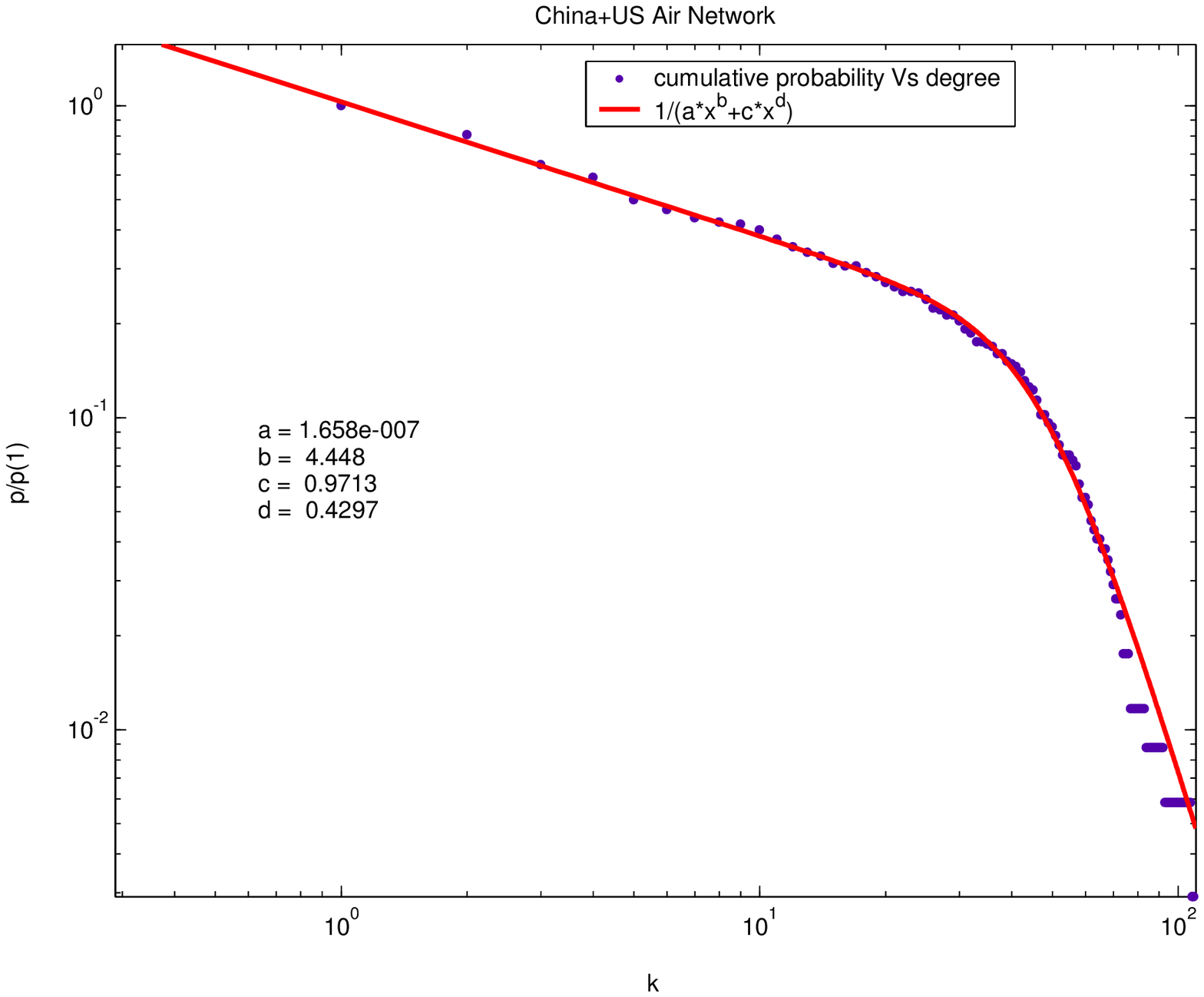}
\caption{Degree distribution (black points) of China+US air
network. The red line is the least squares fitting with Eq.
(\ref{ReverseComposite}).}
\end{figure}


\begin{thebibliography}{99}
\bibitem{QuoteUSAir} Here we only include airports which produce
commercial flights. Actually there are in total around 5000
airports in US, but most of them are either private or of military
use.)
\end{thebibliography}
\end{document}